\documentclass{aa}  

\usepackage{diagbox}
\usepackage{graphicx}
\usepackage{txfonts}
\usepackage[colorlinks=true, linkcolor=blue, citecolor=blue, filecolor=blue, urlcolor=blue]{hyperref} 

\newcommand{\YS}{{\rm YS}}

\newcommand{\rms}{{\rm s}}
\newcommand{\p}{{\rm p}}

\begin{document} 


   \title{The binary Yarkovsky effect on the primary asteroid with applications to singly synchronous binary asteroids}


   \author{Wen-Han Zhou 
    }

   \institute{Universit\'e C\^ote d'Azur, Observatoire de la C\^ote d'Azur, CNRS, Laboratoire Lagrange, 96 Bd de l'Observatoire, Nice 06304, France\\
              \email{wenhan.zhou@oca.eu}
              }

 
  \abstract
    {The binary Yarkovsky effect on the secondary asteroid (BYS) was recently discovered to influence binary asteroid systems by pushing the secondary asteroid towards a synchronous orbit on a short timescale. However, the binary Yarkovsky effect on the primary (BYP) remains less understood, partly due to non-linear effects from partial eclipses, but could have significant implications for singly synchronous binaries. }
{We aim to obtain an empirical formula for the BYP and estimate its induced orbital drifting rates for real binary asteroids.}
{We solved the radiation forces numerically.\ By fitting the numerical results, we find an empirical modified solution to estimate the effective BYP: the traditional BYP formula multiplied by $(r_\rms / r_\p)^{(\alpha -1 )}$, which accounts for the partial eclipse.}
{We confirm that the BYP pushes the primary towards a synchronous orbit where its spin equals the mean motion. Numerical results indicate that the parameter $\alpha$ is relatively insensitive to the ratio of the spin rate to the mean motion and decreases slightly with increasing thermal inertia. For small binary systems with a typical thermal inertia of 200 tiu, $\alpha$ is approximately 1.7. The BYP is found to affect the mutual orbit of singly synchronous binaries with a timescale typically an order of magnitude longer than that of the BYS. Drift rates induced by the BYP for known small binary asteroids (primary radius $< 1$ km) range from -0.001 to -1 cm yr$^{-1}$. A comparative analysis with observed orbital drift rates shows agreement for pre-impact Didymos and 1996 FG$_3$ but discrepancies for 2001 SL$_9$ and 1999 KW$_4$, suggesting complex dynamics in these systems involving the BYP, the binary Yarkovsky-O'Keefe-Radzievskii-Paddack (BYORP) effect, and tides.}
{The BYP is changing the mutual orbits of most discovered binary asteroids. We suggest that the BYP should be considered along with BYORP and tidal effects when studying binary systems' long-term dynamics.}
    
    %
    
   \keywords{minor planets, asteroids: general}

   \titlerunning{Binary Yarkovsky on the primary asteroid}

   \maketitle

%

\section{Introduction}
\label{sec:intro}

Binary asteroids are estimated to account for approximately 15\% of the asteroid population \citep{Margot2015, Virkki2022, Minker2023, Liberato2024}. The dynamical lifetime of near-Earth asteroids and the collisional lifetime of main-belt asteroids are typically comparable to or longer than 10 million years \citep{Farinella1998, Gladman2000}. Understanding the long-term dynamics of binary asteroids is crucial for unravelling their evolutionary paths to their current configurations and deciphering their history.

In classical theory, the tidal effect and the binary Yarkovsky-O'Keefe-Radzievskii-Paddack (BYORP) effect governs the long-term evolution of binary asteroid systems \citep{Cuk2005, Cuk2007, Cuk2010, Jacobson2011b, Jacobson2014}. The tidal effect results from energy dissipation due to deformation, which drives the spin of the object towards a synchronous state and moves the orbit out of a synchronous state \citep{Murray1999}. A synchronous state occurs when the spin period equals the mutual orbital period. The BYORP effect is a net radiative force with a random direction and magnitude averaged over a period and is caused by anisotropic radiation from the asteroid's irregular shape \citep{Cuk2005}. 

However, some aspects are not satisfactorily explained by classical theory. The tidal synchronization timescale for small binaries is considerably long (e.g. $\sim 10~$Myr) unless a small tidal parameter, $Q/k$, is assumed, where $k$ is the Love number and $Q$ is the quality factor \citep{Murray1999, Cuk2005,Quillen2022, Zhou2024b}. The Yarkovsky-O'Keefe-Radzievskii-Paddack (YORP) effect predicts that half of secondaries are synchronous and the other half asynchronous; however, in reality the vast majority of known binaries  (i.e.,$\sim$ 90\%) are singly synchronous where only the secondary is synchronous, like the Earth-Moon system \citep{Zhou2024b}. Furthermore, four binary asteroids (namely the 1996 FG$_3$, 1999~KW$_4$, 2001~SL$_9$, and Didymos systems) have been detected with secular mutual orbital drifts. {BYORP and tidal effects could potentially explain these drift rates, but refinements to shape models and tidal parameters are needed} \citep{Scheirich2015, Scheirich2021, Scheirich2022, Scheirich2024, Naidu2024, Cueva2024, Richardson2024}. This suggests that additional factors may be influencing the long-term dynamics of binary asteroids.

Recently, \citet{Zhou2024b} revisited the Yarkovsky effect on binary asteroid systems, building on pioneering work related to Earth's satellites and planetary rings \citep{Rubincam1982, Milani1987, Farinella1996, Metris1997, Vokrouhlicky2005, Rubincam2006, Vokrouhlicky2007, Rubincam2014}. The binary Yarkovsky effect arises from eclipse-induced thermal perturbations and thermal radiation from the other component in the system. By reorganizing the solution provided by \citet{Vokrouhlicky2007}, \citet{Zhou2024b} found that the binary Yarkovsky effect on a secondary asteroid with a low inclination can drive the mutual orbit towards a synchronous state. The typical timescale for this synchronization process is approximately 0.1 Myr, which can be shorter than the timescales associated with tidal effects and the YORP effect. Thus, it has been proposed that the binary Yarkovsky effect may account for the synchronous state observed in the majority of known binary asteroids \citep{Zhou2024b}. The binary Yarkovsky effect ceases to operate on the secondary once it enters a synchronous state.

However, there should also be a mirror binary Yarkovsky effect on the primary asteroid (hereafter the BYP) similar to the binary Yarkovsky effect on the secondary asteroid (BYS). The BYP is expected to continue modifying the mutual orbit after the BYS is inactive for the synchronized secondary, suggesting that most observed binary asteroids  ($\sim 90\%$) should be influenced by the BYP in addition to tidal effects and the BYORP effect. Preliminary estimates indicate that the orbital drift rate caused by the BYP is weaker than that of the BYS, by roughly a factor of $(r_\rms / r_\p)^2$, where $r_\rms$ is the radius of the secondary and $r_\p$ is the radius of the primary. However, because the primary experiences only a partial eclipse due to the smaller size of the secondary, unknown non-linear effects may be at play.

This study explores the behaviours of the BYP and its influence on singly synchronous binary asteroids and evaluates its typical strength using numerical methods. This paper is organized as follows: Section~\ref{sec:overview} overviews the mechanism and main equations of the binary Yarkovsky effect; Sect.~\ref{sec:numerical} describes the numerical method used in this work to investigate the BYP; Sect.~\ref{sec:results} discusses the numerical result, develops the empirical approximate formula for the BYP, and presents the estimated orbit drift rates of known synchronous binaries due to the BYP.

\begin{figure*}
    \centering
    \includegraphics[width=\linewidth]{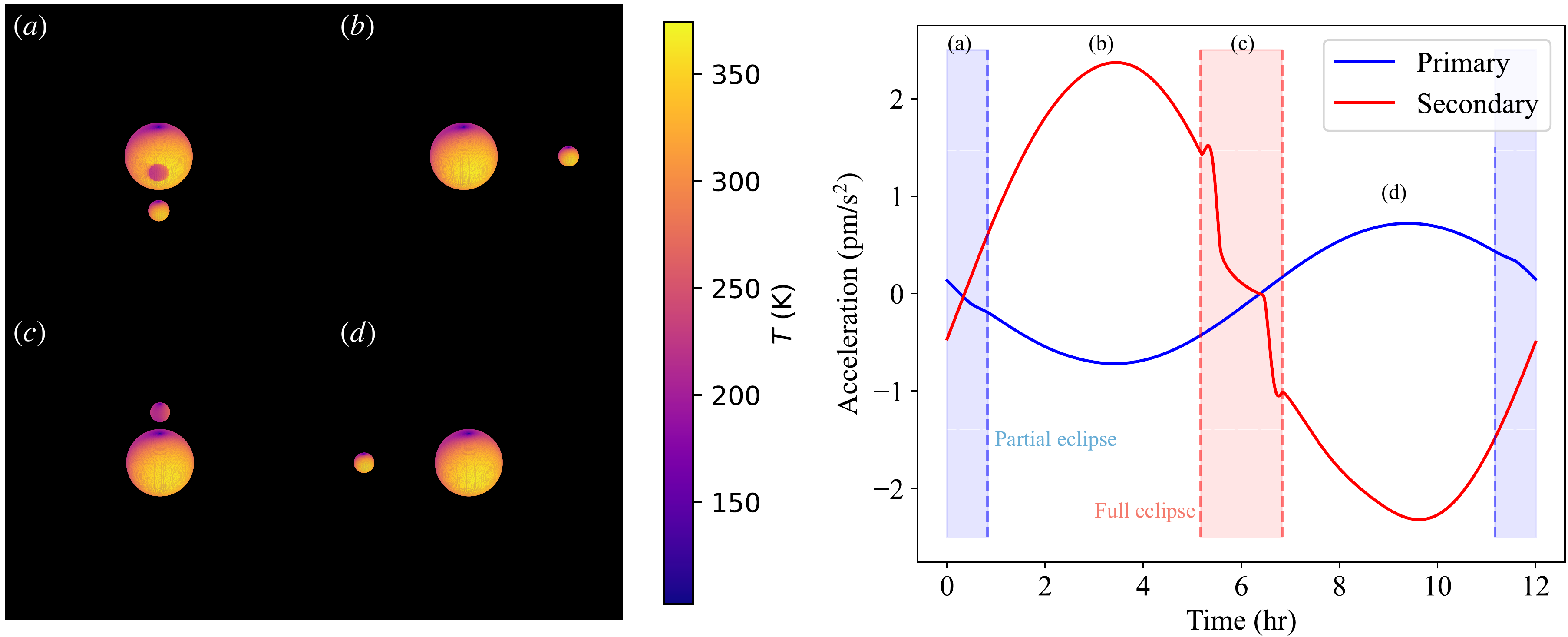}
    \caption{Left: Snapshots of the temperature field of the binary asteroids. Diagrams (a)-(d) illustrate the anti-clockwise orbit of the secondary asteroid around the primary. Both the primary and secondary have spin rates of 3 hours. The other properties of the binary system are detailed in Sect.~\ref{sec:frequency}. In phase (a) the primary is partially eclipsed by the secondary, while in phase (c) the primary fully eclipses the secondary. Right: Tangential accelerations due to thermal forces for the primary (blue) and secondary (red). The eclipse periods are represented by shaded areas.}
    \label{fig:snapshot}
\end{figure*}




\section{Analytical consideration}
\label{sec:overview}
In this paper, subscripts `s' and `p' denote parameters related to the secondary and primary, respectively. According to \citet{Zhou2024b}, for binary asteroids with aligned spin vectors, mutual orbital vectors, and heliocentric orbits, the BYS-induced orbital drift rate is
\begin{equation}
\label{eq:a_dot_YK_s}
    \dot a_{{\rm BY},\rms} =\frac{2 f_{\rm BY,s} \mathcal{F_\rms}}{n}
,\end{equation}
with
\begin{equation}
\label{eq:f_BYS}
    f_{\rm BY,s} = f_{\rm YS, \rms} \left(1 - {\pi r_\p \over 4 a}\right).
\end{equation}
{Here $f_{\rm BY}$ and $f_{\rm YS}$ are the binary Yarkovsky coefficient and the Yarkovsky-Schach (YS) coefficient, respectively.} The $n$ is the mean motion, and the $a$ is the semi-major axis of the mutual orbit. The nominal radiation pressure per unit mass, $\mathcal{F}$, is defined as
\begin{equation}
\label{eq:F}
    \mathcal{F} = \frac{\Phi (1 - A)\,\pi r^2}{m c}.
\end{equation}

The coefficient $f_\YS$ is a complicated function of the physical properties of the components. However, in the limit that the size of the binary components is larger than a few centimetres,  which is true in the context of binary asteroids, $f_\YS$ can be approximated as\begin{equation}
\label{eq:f_YS_s}
    f_{\YS,\rms} = \frac{4r_\p}{9\pi a} {\Theta_{\rms} \over 2 + 2\Theta_{\rms} + \Theta_{\rms}^2}\, {\rm Sign}(n - \omega_\rms)
.\end{equation}
{Here $\omega$ is the spin rate, and $r_{\rm p}/\pi a$ accounts for the time fraction of the eclipse.} The thermal parameter, $\Theta_{\rms}$, is defined as
\begin{equation}
\label{eq:Theta_rms}
    \Theta_\rms = {\Gamma \sqrt{|\Delta|} \over \varepsilon \sigma T_{\rm sub}^3}. 
\end{equation}
Here $\Gamma$ is the thermal inertia, $\Delta_\rms = \omega_\rms - n$ is the relative frequency, $\epsilon$ is the emissivity, $\sigma = 5.67 \times 10^{-8} \rm W~m^{-2}~k^{-4}$ is the Stefan-Boltzmann constant, and $T_{\rm sub} = ((1-A)\Phi/\epsilon \sigma)^{1/4} $ is the subsolar temperature.

The total binary Yarkovsky effect is the combined contribution of both the primary and secondary asteroids. For the primary component, in principle, there is a mirror effect to the secondary. The formula for the primary can be obtained by simply replacing the subscript `s' with `p' in Eqs.~\ref{eq:a_dot_YK_s} and \ref{eq:f_YS_s}, which gives
\begin{equation}
\label{eq:f_YS_p}
    f_{\YS,\p} = \frac{4r_\rms}{9\pi a} {\Theta_{\p} \over 2 + 2\Theta_{\p} + \Theta_{\p}^2}\, {\rm Sign}(n - \omega_\p).
\end{equation}

However, due to the complexity of the geometry of the shadow cast on the primary component by the secondary component, this may not be accurate. In the derivation of Eq.~\ref{eq:f_YS_s}, we assumed that the entire body stops receiving sunlight once its mass centre enters the shadow region, for the sake of simplifying the calculation. This assumption is reasonable for the secondary but not for the primary component since only a portion of the primary enters the eclipse at any given time (similar to a solar eclipse on Earth caused by the Moon). This partial eclipse reduces the strength of the YS effect, which heavily depends on the shadow; because of its induced non-linearity, a numerical approach is required to solve it. 

This study ignores {the planetary Yarkovsky (pY) effect on the primary, which is caused by the radiation from the secondary asteroid, as it is expected to be minimal}. It would only induce a correction factor of $(1 - \pi r_\rms / 4 a ) \sim 1$ for the overall effect for the primary.

\section{Numerical model}
\label{sec:numerical}

The radiation force of a surface element is $\propto \sigma T^4 S$, where $T$ denotes the temperature and $S$ denotes the surface area. Thus, to obtain the radiation force, we need to obtain the surface temperature of the shape model. 

An irregular shape could produce the BYORP torque \citep{Cuk2005}. To eliminate the influence of the BYORP effect, we chose a spherical shape model approximated by a polyhedron with 1280 triangulated facets as the shape model. {This symmetric shape model introduces a BYORP-induced error of only $10^{-6}$ to $f_{\rm YS}$, which is negligible compared to $f_{\rm YS} \sim 10^{-3}$.}

The temperature, $T,$ of the surface and the layer beneath is governed by 
\begin{equation}
\label{eq:heat_diffusion}
    \frac{\partial T}{\partial t} = \frac{\kappa}{C \rho } \frac{\partial^2 T}{ \partial z^2},
\end{equation}
with two boundary conditions, 
\begin{align}
    & \kappa \frac{\partial T }{\partial z} \left \vert_{z = 0} = E(t) - e \sigma T^4 \vert_{z=0} \label{eq:BC1}  \right. ,\\
    & \kappa \frac{\partial T}{\partial z} \left \vert_{z \to \infty} = 0, \right. \label{eq:BC2}
\end{align}
where $t$ is the time, $\kappa$ is the thermal conductivity, $C$ is the specific heat capacity, $\rho$ is the bulk density of the asteroid, and $e$ is the emissivity. In this study we set $\kappa = 0.1$~W~m$^{-1}$~K$^{-1}$, $C = 550$~J~K$^{-1}$~kg$^{-1}$, and $\rho = 2000$~kg~m$^{-3}$ for both the primary and secondary components by default. This gives a thermal inertia $\Gamma = \sqrt{C \rho \kappa} \sim 330 \rm ~tiu$ [$\rm J~m^{-2}~K^{-1}~s^{-1/2}$], which is close to the measured mean value of $\sim 200\rm~ tiu$ for kilometre-sized near-Earth objects \citep[NEOs;][]{Delbo2007}. {According to in situ measurements, asteroids Ryugu and Bennu have thermal inertia values of $225 \pm 45$ tiu \citep{Shimaki2020} and $350 \pm 20$ tiu \citep{Dellagiustina2019}, respectively.} The one-dimensional heat conduction equation (Eq.~\ref{eq:heat_diffusion}) is sufficient for our problem since our interests lie in binary asteroid systems whose components are much larger than the thermal penetration depth; the latter is approximately a few centimetres. A full three-dimensional heat conduction equation is needed when investigating small (sub-centimetre-scale) dust.
 
We solved the temperature $T_{i,j,k}$ at a depth of $(j-1)\delta z$ below the $i$-th facet at the $k$-th time step using the numerical scheme described in \citet{Zhou2024}. The thermal equilibrium state, where the intake energy equals the released energy of the system, is usually established after $\sim 50$ mutual orbits unless the thermal inertia is considerably high (e.g. > 1000 tiu). A snapshot of our simulation is shown in Fig.~\ref{fig:snapshot}. With the surface temperature, one can obtain the radiation force in the $k$-th timestep for the whole body via\begin{equation}
    \vec F_{k} = -\frac{2\epsilon \sigma}{3c}\sum_i  T_{i, 0, k}^4 S_{ i} \vec n_{i},
\end{equation}
{where $S_i$ and $\vec n_{i}$  are the area and the normal vector of the $i$-th surface element, respectively.} The tangential component of the force, which is the source of the Yarkovsky effect, in the $k$-th timestep is then easily obtained as $\vec F_{k} \cdot \vec v_{ k} / v_{k}$, assuming a circular orbit. The YS coefficient is calculated as\begin{equation}
    f_{\rm YS} = \frac{1}{k_{\rm max}} \sum_k \frac{\vec F_{ k} \cdot \vec v_{k} \cdot c}{ v_{ k} \Phi (1-A) \sum_i S_{ i}  },
\end{equation}
{with $A$, the albedo of the asteroid, being $\sim 0.1$.} The mutual heating between the primary and the secondary was ignored due to its minimal contribution to the overall effect on the primary.

\section{Results and implications}
\label{sec:results}

We investigate the YS coefficient as a function of the frequency ratio, $\omega_{\rm p}/n,$ in Sect.~\ref{sec:frequency} and propose an empirical modified formula for BYP in Sect.~\ref{sec:size}. Finally, we discuss the implications of the BYP on the long-term evolution of binary asteroids and estimate its magnitude for known synchronous binary asteroids in Sect.~\ref{sec:discussion}.

For simplicity, we assumed the spin vector of the primary asteroid aligns with the mutual orbit vector and considered the mutual orbit to be  co-planar with the heliocentric orbit, {given the low inclinations (i.e. $i < 30^\circ$) observed in real binary asteroids \citep{Pravec2012}}. Low inclinations are likely common as these systems are believed to form via YORP-driven spin-up and mass shedding followed by re-accumulation in the equatorial plane of the primary \citep{Walsh2008, Pravec2012, Zhang2021, Agrusa2024}. {We note that this assumption leads to an upper limit for our estimated Yarkovsky effect. The dependence of the BYS and BYP on non-zero inclinations is discussed in Appendix~\ref{AppA}.}

\subsection{Synchronization of the mutual orbit}
\label{sec:frequency}

We considered a typical small binary asteroid system on a heliocentric circular orbit with a heliocentric semi-major axis $a_{\rm h}=1$~au. We set the semi-major axis of the mutual $a \simeq 3.1 r_\p$ to keep the orbital period $P = 12~$h. By varying the spin periods of the primary and the secondary, we obtain the YS coefficient as a function of the frequency ratio, $\omega / n,$ which can be easily translated to $\Delta_\rms$ via Eq.~\ref{eq:f_YS_s} and $\Delta_\p$ via Eq.~\ref{eq:f_YS_p}. 

\begin{figure}
    \centering
    \includegraphics[width=0.5\textwidth]{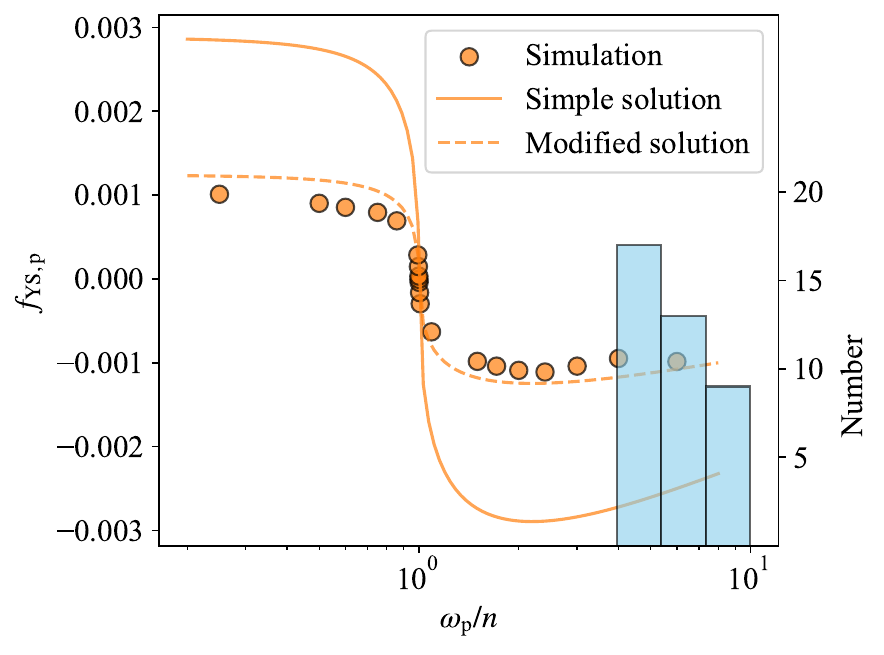}
    \caption{YS coefficient for the primary as a function of the ratio of spin velocity to mean motion. The primary asteroid has a radius of 1 km, and the secondary has a radius of 0.3 km. Gold circles represent the numerical results. The solid gold curve shows the simple analytical solution from Eq.~\ref{eq:f_YS_p}, while the dashed gold curve shows the modified solution from Eq.~\ref{eq:f_YS_fit2} with $\alpha = 1.7$. The $\omega_\p/n$ distribution of confirmed binary asteroids is depicted in the blue histogram, which is truncated at 10.}
    \label{fig:YS_m}
\end{figure}

Figure~\ref{fig:YS_m} shows the YS coefficient, $f_{\rm YS, p}$, for the primary, obtained from the numerical model and the simple analytical model, as a function of the frequency ratio, $\omega_\p/n$. The Yarkovsky force has the same sign as the YS coefficient. The YS coefficient is positive when $\omega_\p / n$ is smaller than 1, indicating the mutual orbit will expand under the Yarkovsky force. On the other hand, when $\omega_\p /n$ is larger than 1, the YS coefficient is negative, implying a shrinking orbit. The Yarkovsky force vanishes when the primary is in a synchronous state, $\omega_\p = n$. Therefore, we conclude that the BYP pushed the mutual orbit towards the synchronous orbit of the primary component. 

However, we observe that the magnitude of the numerical results deviates slightly from the analytical results, which may be due to the partial eclipse on the primary, as discussed in the following section.

\subsection{Empirical formula for the primary}
\label{sec:size}

\begin{figure*}
    \centering
    \includegraphics[width=\textwidth]{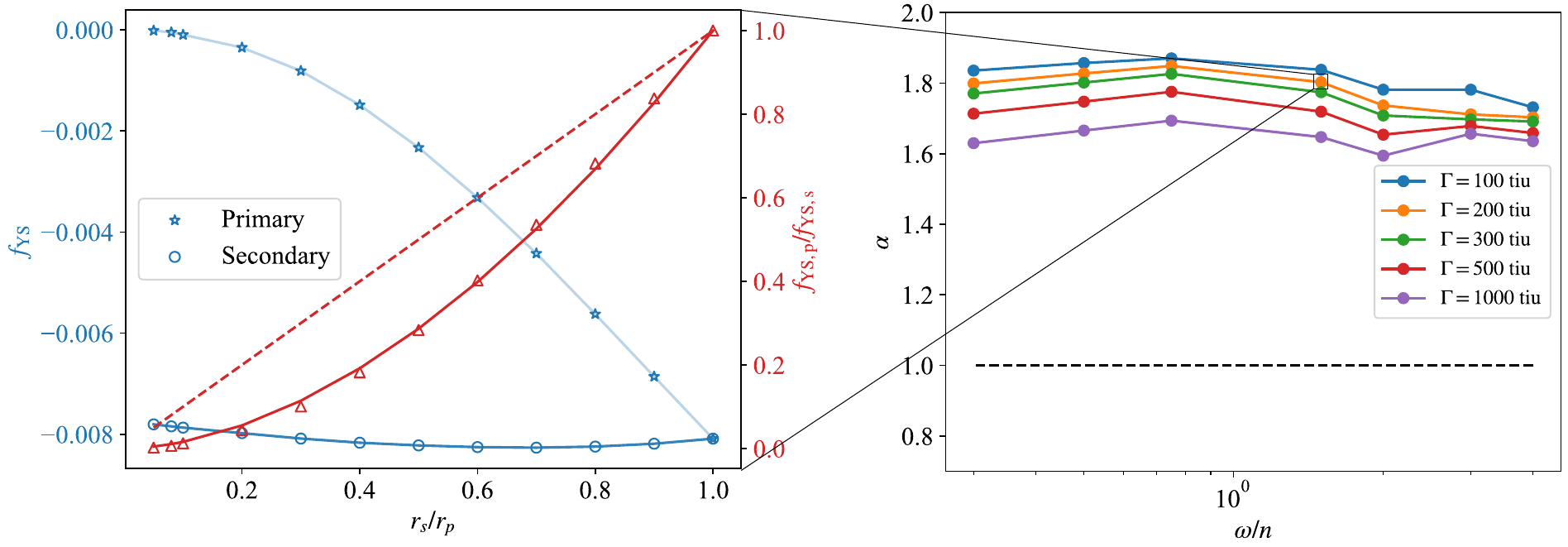}
    \caption{Left: YS coefficient as a function of the secondary-to-primary size ratio. The primary asteroid has a radius of 1 km. The orbital period is 12 hours, and the spin periods of both components are 8 hours. The ratio of the primary-to-secondary YS coefficient is indicated by red triangles. The solid red line represents the best-fit curve: $f_{\rm YS,p}/f_{\rm YS,s} = (r_{\rm s} / r_\p)^{1.80}$. The dashed red line shows the theoretical formula without accounting for the non-linear effect of partial eclipses: $f_{\rm YS,p}/f_{\rm YS,s} = (r_{\rm s} / r_\p)$. Right: Value of $\alpha$ as a function of the frequency ratio and thermal inertia. The dashed black line represents the theoretical value, excluding the non-linear effects of partial eclipses. For a typical small binary system with $\Gamma \sim 200$ tiu and a primary spin period of approximately three hours, $\alpha$ is around 1.8.}
    \label{fig:YS_size}
\end{figure*}

Ideally, the ratio of $f_{\rm YS,p}$ to $f_{\rm YS,s}$ should be proportional to $r_\rms / r_\p$. However, due to the non-linear effect produced by the partial eclipse stated above, the YS coefficient for the primary, assuming the same physical properties as the secondary except the size, is affected by a non-linear dependence on $r_\rms $. As a result, the ratio of $f_{\rm YS,p}$ to $f_{\rm YS,s}$ is better described as 
\begin{equation}
\label{eq:f_YS_fit2}
    \Tilde{f_{\YS,\p}} = \left(\frac{r_\rms}{r_\p} \right)^{\alpha} f_{\YS,\rms} = \left(\frac{r_\rms}{r_\p} \right)^{\alpha -1} f_{\YS,\p}
,\end{equation}
where $f_{\YS, \rms}$ and $f_{\YS,\p}$ can be obtained from Eqs.~\ref{eq:f_YS_s} and \ref{eq:f_YS_p}, respectively.


We searched the best-fit $\alpha$ for the numerical results by minimizing the sum of the squares of the difference between simulation results and the fitted values, as shown in the left panel in Fig.~\ref{fig:YS_size}. The right panel of Fig.~\ref{fig:YS_size} shows the values of $\alpha$ under different frequency ratios, $\omega/n$, and thermal inertia.\ We find that $\alpha$ is insensitive to $\omega/n$ but shows a clear decreasing trend with increasing thermal inertia. The values of $\alpha$ are $1.81 \pm 0.05$, $1.77 \pm 0.05$, $1.75 \pm 0.05$, $1.70 \pm 0.04$, and $1.65 \pm 0.03$ for $\Gamma = 100,~200,~300,~500$, and $1000$~tiu [$\rm J~m^{-2}~s^{-1/2}~K^{-1}$], respectively. One-kilometre-sized objects typically have a thermal inertia of around 300 tiu \citep{Delbo2007}, as seen in Didymos \citep{Rivkin2023, Rozitis2024}. Once $\alpha$ was determined, we obtained the modified $f_{\rm YS,p}$ solution. An example of the modified $f_{\rm YS,p}$ solution with $\alpha = 1.7$ is shown in Fig.~\ref{fig:YS_m} (dashed gold curve). It fits the simulation much better than the simple solution with $\alpha = 1$.




\subsection{Implications}
\label{sec:discussion}

\begin{figure*}
    \centering
    \includegraphics[width=\linewidth]{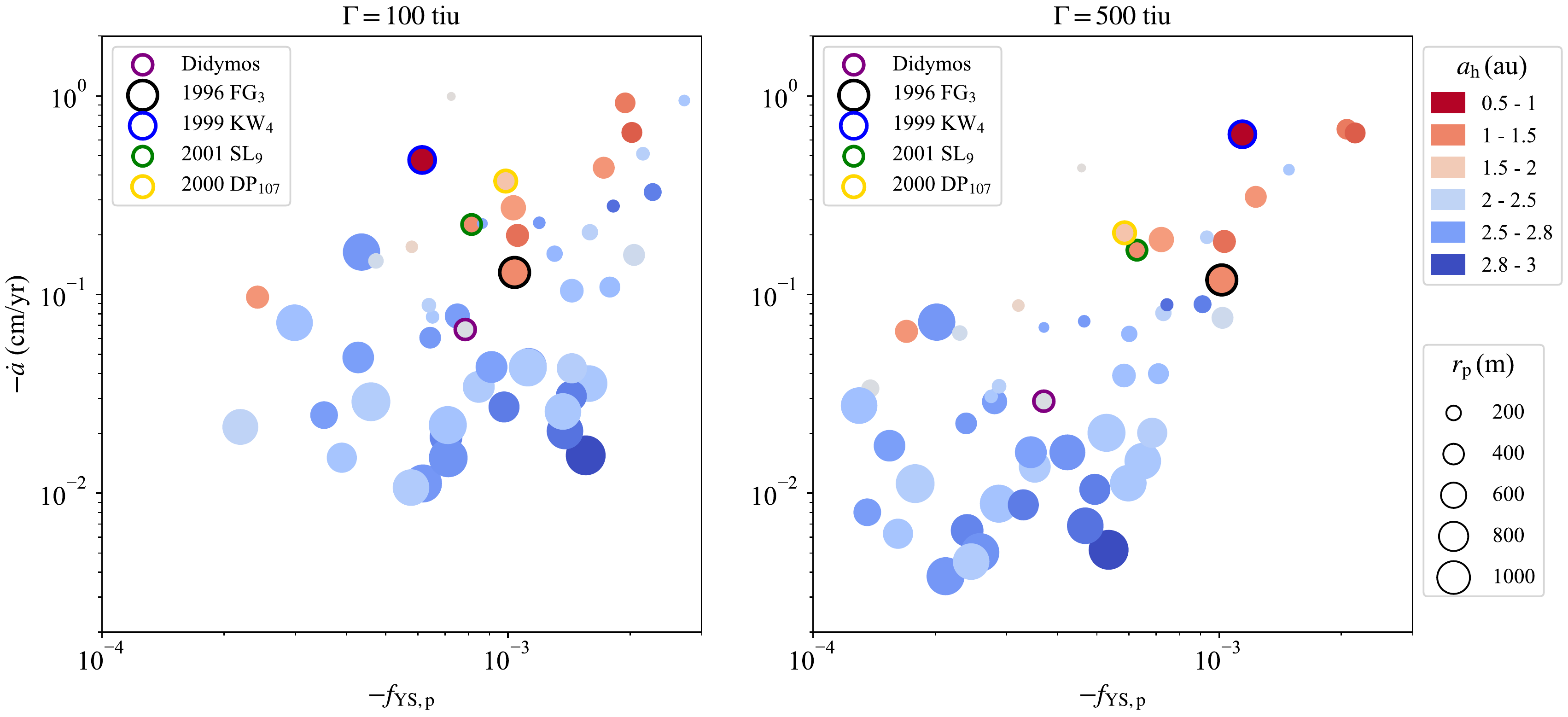}
    \caption{YS coefficient and YS-induced orbital drift rate for confirmed small binary asteroids, assuming thermal inertia of 100 tiu (left) and 500 tiu (right). The colours indicate the heliocentric semi-major axis, with bluer colours representing greater distances from the Sun. The size of each circle corresponds to the size of the primary asteroid. As thermal inertia increases, blue dots tend to move downwards and red dots upwards, although the outcome is a complex function of thermal inertia.}
    \label{fig:asteroids}
\end{figure*}

We have demonstrated that the BYP moves the primary asteroid towards a synchronous orbit, albeit on a longer timescale than the BYS. Together with the BYS \citep{Zhou2024b} and the BYORP effect \citep{Cuk2005}, this forms a unified description of radiative forces acting on binary asteroid systems, as summarized in Table~\ref{tab1}: For an asynchronous component, the binary Yarkovsky effect is at play; for a synchronous component with an asymmetric shape, the binary YORP effect is dominant; and for a synchronous component with a symmetric shape, there is no significant secular radiative effect. 

\renewcommand{\arraystretch}{1.5} 
\begin{table}[h]
    \centering
    \caption{Radiative effects on {the object in a binary system}.}
    \label{tab1}
    \begin{tabular}{|c|c|c|} 
        \hline
        \diagbox[width=2.4cm]{Shape}{Orbit} & {Synchronous} & {Asynchronous} \\ 
        \hline 
        {Symmetric} & Null & Binary Yarkovsky \\ 
        \hline
        {Asymmetric} & Binary YORP & Binary Yarkovsky \\ 
        \hline
    \end{tabular}
    \tablefoot{The binary Yarkovsky effect includes the pY effect and the YS (eclipse-induced) effect.}
\end{table}

Based on this framework, we summarize the primary dynamics for different types of binary asteroids as follows. For synchronous binaries where both components are asynchronous, the BYS, the BYP, and tidal effects are all active. The BYS dominates over the BYP and, with the assistance of tidal effects, drives the secondary component towards a synchronous state. The synchronization timescale for a binary asteroid system with a primary radius of approximately 1 km is roughly 0.1 million years \citep{Zhou2024b}. For larger objects, the tidal effect predominates over the BYS. The YORP effect may also contribute to synchronization, though its stochastic nature is not yet well constrained \citep{Statler2009, Golubov12, Bottke2015, Zhou2022}.

For singly synchronous binaries where the secondary is synchronous, which is the most common type of known binary asteroids,  BYP, BYORP (acting on the secondary), and tidal effects are all active. The BYP is expected to shrink the orbit of most binary asteroids given that the majority of primary asteroids in the main belt or in near-Earth orbit rotate rapidly, close to the spin limit of 2.2 hours \citep{WalshJacobson2015}. The timescale of the BYP is typically an order of magnitude longer than that of the BYS and depends on the size ratio of the two components. The BYORP effect is predicted to act on a timescale ranging from 0.01 to 10 million years with a random direction and depends on factors such as shape \citep{Cuk2005, Mcmahon2010,Steinberg2011, Jacobson2011b}, surface roughness \citep{Cuk2024}, and rotational state \citep{Quillen2022}.

For doubly synchronous binaries (i.e. both components are synchronous), the BYORP effect acts on both components. Tidal effects also become significant in eccentric orbits \citep{Wisdom2008, Goldberg2024} or for librating objects \citep{Scheeres2009, Jacobson2014}.

While the magnitude of the BYORP effect remains largely uncertain due to the unconstrained shape and fine surface structures of most asteroids, the BYP can be estimated quickly using the formula in this work (see Eq.~\ref{eq:f_YS_fit2}). We have estimated the YS coefficients and orbital drift rates for known synchronous binary asteroids with detected spin periods and orbital periods \citep{Pravec2007, Warner2009, Pravec2012, Pravec2016, Monteiro2023}. As an order-of-magnitude approximation, we assumed spherical binaries that have relatively low inclinations, which ensures that these binaries experience eclipses during each mutual orbit. This assumption is appropriate for most binary asteroids \citep{Pravec2012}. {The binary Yarkovsky effect is expected to be insensitive to shape, except in cases of extreme irregularity, which is unrealistic for binary asteroids. These bodies are expected to have relatively regular shapes following rotational disruption.} All of these binary asteroids shrink the mutual orbit under the BYP due to their faster spin relative to the mean motion (see also the histogram in Fig.~\ref{fig:YS_m}).

Figure~\ref{fig:asteroids} shows that for the known small binary asteroids, the absolute value of $f_{\rm YS,p}$ ranges from $10^{-4} $ to a few $\times 10^{-3}$ and the BYP-induced orbital drifting rate from $-0.001~\rm cm~yr^{-1}$ to $-1~\rm cm~yr^{-1}$. {For comparison, the typical BYORP coefficient is $\sim 10^{-3}$ \citep{Mcmahon2010,Jacobson2011b} and decreases with reduced relative surface roughness \citep{Cuk2024}.} There is a clear trend that larger and colder binary asteroids have smaller values of $f_{\rm YS,p}$ and $\dot a$ (located towards the left and bottom of the plot). Thermal inertia influences these results in a complex manner. Generally, within the range of 100 to 500 tiu, as thermal inertia increases, the Yarkovsky effect decreases for more distant objects (e.g. middle and outer main belt objects), while it strengthens for closer objects (e.g. NEOs and inner main belt objects).

We compared our results with four binary asteroids that have observed orbital migration. The estimated binary Yarkovsky effect versus observed values are as follows: -0.05 versus $-0.08 \pm 0.02$~cm~yr$^{-1}$ \citep{Scheirich2022, Scheirich2024, Naidu2024} for pre-impact Didymos, -0.079 versus $-0.07 \pm 0.34$~cm~yr$^{-1}$ \citep{Scheirich2015} for 1996 FG$_3$, -0.13 versus $-2.8\pm 0.2$~cm~yr$^{-1}$ \citep{Scheirich2021} for 2001 SL$_9$, and -0.19 versus 1.2~cm~yr$^{-1}$ \citep{Scheirich2021} for 1999 KW$_4$. For pre-impact Didymos and 1996 FG$_3$, the BYP predicts orbital drift rates close to the observed values. However, for asteroid 2001 SL$_9$, our estimated BYP-induced orbital drift rate is an order of magnitude lower than the observed value. Furthermore, the outward drift of 1999 KW$_4$ cannot be explained by the BYP alone, suggesting the presence of additional mechanisms, such as a strong tidal effect, a BYORP effect on a synchronous secondary, or the BYS on an asynchronous secondary. Further observational data are needed to better understand the complex long-term dynamics of binary asteroids.

In \citet{Vokrouhlicky2005}'s pioneering work on the Yarkovsky effect in binary asteroids, simulations of asteroid 2000 DP$_{107}$ showed that the mean transverse acceleration of the mutual orbital motion is $\sim  - 6 \times 10^{-15}~\rm m~s^{-2}$ assuming $K = 0.1 ~\rm W~m^{-1}~K^{-1}$, $C = 800~\rm J~kg^{-1}~K^{-1}$, and $\rho = 1.7~\rm g~cm^{-3}$ for both components. This translates to an orbital drift rate of about $\dot a \sim - 0.85 \rm cm~yr^{-1}$. In their simulation, the primary and secondary are both spherical polyhedral shape models with the assumption of a synchronous secondary. Therefore, their results reflect the BYP. Our theoretical estimate of $-0.27 \rm ~cm~yr^{-1}$ is consistent in both sign and order of magnitude with the previous simulation result despite some deviation that may arise from the system's eccentricity.


\section{Conclusion}

This work, together with our previous work on the BYS \citep{Zhou2024b}, completes the basic theoretical framework of the binary Yarkovsky effect. The BYP is shown to modify the mutual orbit after the secondary asteroid reaches synchronization. The timescale of the BYP is generally an order of magnitude longer than that of the BYS and depends on the size ratio between the secondary and primary. 

We propose an empirical modified formula to estimate the BYP: applying the traditional binary Yarkovsky formula (Eq.~\ref{eq:f_YS_p}) and then multiplying it by $(r_\rms / r_\p)^{(\alpha - 1)}$ (see Eq.~\ref{eq:f_YS_fit2}). Our numerical results indicate that $\alpha$ is relatively insensitive to the frequency ratio, $\omega / n$, but decreases with increasing thermal inertia. For a typical small binary asteroid system with $\Gamma = 200$ tiu, the best-fit value for $\alpha$ is approximately 1.7. We summarize the primary mechanisms for binary asteroid systems as follows: the BYS and tidal effects are active for asynchronous binaries, while BYS, BYORP, and tidal effects operate in singly synchronous binaries, which are the most commonly observed binary asteroids.

We estimated BYP-induced drift rates for known small binary asteroids with primary radii $r_\p < 1$ km. These drift rates range from -0.001 to -1 cm yr$^{-1}$. For middle and outer main belt objects, the BYP tends to decrease with increasing thermal inertia, whereas for NEOs and inner main belt objects, the BYP increases with increasing thermal inertia.

We compared our results with the observed orbital drift rates of four binary systems. Our findings are consistent with the pre-impact Didymos and 1996 FG$_3$ systems but show discrepancies for the 2001 SL$_9$ and 1999 KW$_4$ systems. We suggest that a complicated model involving BYP, BYORP, and tidal effects is required to fully understand the long-term dynamics of binary systems.

\begin{acknowledgements}
I thank the referee for the constructive suggestions. I thank David Vokrouhlick\'y for his valuable suggestions. I am also grateful to Patrick Michel and Seiji Sugita for their support of this research. I would like to acknowledge the mobility aid from the Universit\'e C\^ote d'Azur and funding support from the Chinese Scholarship Council (No. 202110320014).
\end{acknowledgements}

\bibliography{references}
\bibliographystyle{aa}

\appendix
\section{Discussion on non-zero inclination cases}
\label{AppA}
The inclination $i$ is defined as the angle between the vectors of the mutual orbit and the heliocentric orbit. A non-zero inclination complicates the thermal perturbation caused by the shadow. We offer an estimate of the non-zero inclination and justify the importance of BYP on real binary asteroids. A comprehensive numerical investigation on how a non-zero inclination affects both BYS and BYP is out of the scope of this paper and is left for future study due to its complexity. Our following simple estimation is based on the principle that the strength of the YS effect is generally proportional to the solar flux loss due to the shadow. We will use an inclination-dependent factor $f_{i}$ to describe the ratio of the YS effect for a non-zero inclination relative to the case with zero inclination.

To begin, we considered how inclination influences the YS effect on the secondary. When the inclination is non-zero, the shadow cast by the secondary on the primary no longer remains fixed at the equator but oscillates around it, reaching a maximum altitude, $\beta$. Because the primary is approximately spherical, the duration it experiences shadowing is roughly proportional to $\cos \beta$, leading to a weakening factor of
\begin{equation}
    f_{i,\rm s} = \cos \beta.
\end{equation}
The parameter $\beta$ can be estimated via simple geometry,
\begin{equation}
    \sin \beta = {a \tan i \over r_{\rm p}}.
\end{equation}
A $\sin \beta < 1$ naturally necessitates a low inclination, $i < \arctan (r_{\rm p} / a) \simeq r_{\rm p} / a$, for the occurrence of the shadow.

The case for the primary would be a bit different. The shadow cast by the secondary on the primary is above the equator rather than at the equator when the inclination is not zero. These high-altitude areas affected by the shadow have lower temperatures than the equator, as the total incident solar energy decreases by a factor of $\cos \beta$. Consequently, the overall YS effect would be reduced by a factor of
\begin{equation}
    f_{i, \rm p} = \cos^2 \beta.
\end{equation} 

Recall that the binary Yarkovsky effect also includes the pY effect in addition to the YS effect. The strength of pY effect is approximated by the YS coefficient multiplied by $- \pi r_{\rm p} / a $ for the secondary and $- \pi r_{\rm s} / a $ for the primary. Therefore, compared to Eq.~\ref{eq:f_BYS}, the total binary Yarkovsky coefficient becomes more complicated for non-zero inclination cases: 
\begin{equation}
\label{eq:f_BYS_i}
    f_{\rm BY,s} =  \left( \cos \beta - {\pi r_{\rm p} \over 4 a} \right)  f_{\rm YS, s}
\end{equation}
for the secondary and 
\begin{equation}
\label{eq:f_BYP_i}
    f_{\rm BY,p} =  \left( \cos^2 \beta - {\pi r_{\rm s} \over 4 a} \right)  f_{\rm YS, p}
\end{equation}
for the primary. We note that the typical values for $f_{\rm YS, s}$ and $f_{\rm YS, p}$ are 0.01 \citep{Zhou2024b} and 0.001 (Fig.~\ref{fig:YS_m}), respectively. 

\begin{figure}
    \centering
    \includegraphics[width=\linewidth]{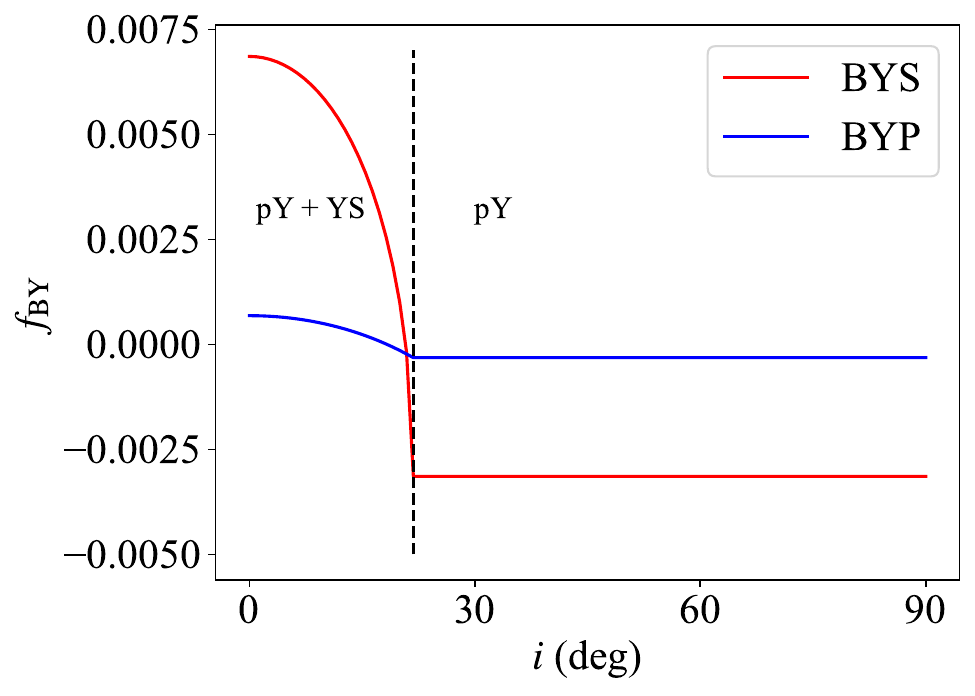}
    \caption{ Binary Yarkovsky coefficient as a function of the inclination for the secondary (red) and the primary (blue). The dashed black line is the critical inclination (i.e. $\arctan (r_{\rm p} / a)$), within which the eclipse always occurs during a mutual orbit. The values of parameters to calculate $f_{\rm BY}$ (Eqs.~\ref{eq:f_BYS_i} and \ref{eq:f_BYP_i}) are set as follows: $a/r_{\rm p} = 2.5$, $f_{\rm BY,s} = 0.01$ and $f_{\rm BY,p} = 0.001$. In this figure, a positive value of $f_{\rm BY}$ represents the migration direction towards the synchronous orbit and a negative value denotes the opposite direction to the synchronous orbit.}
    \label{fig:inclination}
\end{figure}

Figure~\ref{fig:inclination} shows the value of the binary Yarkovsky coefficient as a function of inclination, described by Eqs.~\ref{eq:f_BYS_i} and \ref{eq:f_BYP_i}. It is seen that YS decreases with the inclination and vanishes with a critical inclination, within which the eclipse always occurs during a mutual orbit. In contrast, the pY effect remains independent of inclination, as it is driven by radiation from the companion object within the binary system. There is a specific inclination at which pY cancels out YS, resulting in the disappearance of the binary Yarkovsky effect.

When the inclination is low and YS dominates, the binary Yarkovsky effect tends to synchronize the orbit. Conversely, at high inclinations where pY is dominant, the effect causes de-synchronization. The assumption of zero inclination in the main text maximizes the binary Yarkovsky effect and ensures mutual orbital synchronization. Although the strength of the binary Yarkovsky effect diminishes with increasing inclination, it does not change by more than an order of magnitude except near the critical inclination (i.e. $\arctan (r_{\rm p} / a)$), justifying the importance of the binary Yarkovsky effect for most of the binary asteroids.

We note that a more detailed numerical investigation of the role of the inclination is required in the future, especially considering the non-zero obliquities that complicate the problem significantly.

\end{document}